\documentclass[runningheads]{llncs}

\usepackage{hyperref}
\usepackage[utf8]{inputenc} 
\usepackage[T1]{fontenc}
\usepackage{amssymb}
\usepackage{listings} 
\lstdefinestyle{Muli}{
	language=Java,
	morekeywords={free,fail,solve,getAllSolutions,getAllSolutionsEx,getOneSolution,getOneSolutionEx,muli,search},
	captionpos=b,
	tabsize=2,
	abovecaptionskip={4\p@},
	belowcaptionskip={0\p@},
	showstringspaces=false,
	basicstyle=\ttfamily\footnotesize 
}
\usepackage{cleveref} 
\Crefname{section}{Sect.}{Sections}
\Crefname{listing}{List.}{Listings}
\Crefname{table}{Tab.}{Tables}
\Crefname{figure}{Fig.}{Figures}
\renewcommand{\figurename}{Figure}
\usepackage{microtype}

\usepackage{cleveref}
\usepackage{graphicx} 
\usepackage[format=hang,caption=false]{subfig} 
\usepackage{tikz} 
\usetikzlibrary{calc}
\usetikzlibrary{positioning} 
\usetikzlibrary{arrows} 
\usetikzlibrary{shapes.geometric} 
\usetikzlibrary{matrix,fit} 
\usepackage{tikz-qtree}
\usepackage{tikz-uml}
\usepackage{erciscolors}

\usepackage{booktabs} 

\usepackage{pgfplots}
\def\addlegendimage{\csname pgfplots@addlegendimage\endcsname}
\pgfkeys{/pgfplots/number in legend/.style={%
		/pgfplots/legend image code/.code={%
			\node [anchor=east]at (0.7,-0.0225){#1}; 
		},%
	},
}

\begin{document}
	\renewcommand{\thelstlisting}{\arabic{lstlisting}} 



	%
	%
        \title{Structured Traversal of Search Trees in Constraint-logic Object-oriented Programming}
        \titlerunning{Structured Traversal of Search Trees in Constraint-logic OO Programming}


	\author{Jan C. Dageförde\inst{1}\orcidID{0000-0001-9141-7968}
	\and
	Finn Teegen\inst{2}\orcidID{0000-0002-7905-3804}
	}
	\authorrunning{J. C. Dageförde and F. Teegen}
	\institute{ERCIS, Leonardo-Campus 3, 48149 Münster, Germany\\
		\email{dagefoerde@uni-muenster.de}
		\and
		Institut für Informatik, CAU Kiel, 24098 Kiel, Germany\\
		\email{fte@informatik.uni-kiel.de}
	}

	%

	\maketitle              

	\begin{abstract}
In this paper, we propose an explicit, non-strict representation of search trees in constraint-logic object-oriented programming.
Our search tree representation includes both the non-deterministic and deterministic behaviour during execution of an application.
Introducing such a representation facilitates the use of various search strategies.
In order to demonstrate the applicability of our approach, we incorporate explicit search trees into the virtual machine of the constraint-logic object-oriented programming language Muli.
We then exemplarily implement three search algorithms that traverse the search tree on-demand: depth-first search, breadth-first search, and iterative deepening depth-first search.
In particular, the last two strategies allow for a complete search, which is novel in constraint-logic object-oriented programming and highlights our main contribution.
Finally, we compare the implemented strategies using several benchmarks.

	\end{abstract}

	%
	%


\section{Motivation} \label{sec:intro}

In constraint-logic object-oriented programming, combining imperative code with features from logic programming causes the runtime to execute parts of the imperative code non-deterministically (``don't know'' non-determinism).
To give an example, the program (or search region) depicted in \Cref{lst:simplest-coin} has two solutions. 
The example is written using the Münster Logic-Imperative Language (Muli), which is explained in \Cref{sec:muli}.
The search region declares a boolean logic variable \lstinline|coin|.
Subsequently, evaluating the \lstinline|if| statement causes the runtime environment to take and implement a decision regarding the potential value of \lstinline|coin|, thus introducing non-determinism.
Consequently, implementing the decision selects a single branch of execution, eventually resulting in one of the two outcomes.

\begin{figure}[t]
  \begin{minipage}[b]{0.48\linewidth}
\begin{lstlisting}[caption={A simple non-deterministic search region in Muli for the demonstration of constraint-logic object-oriented programming concepts.}, label={lst:simplest-coin},basicstyle=\ttfamily\footnotesize]
boolean flipCoin() {
	int coin free;
	if (coin == 0)
		return false;
	else
		return true; }
\end{lstlisting}
  \end{minipage}
  \hfill
  \begin{minipage}[b]{0.48\linewidth}
\begin{lstlisting}[caption={Muli search region example that comprises two solutions and a failure.}, label={lst:two-coins}]
boolean flipTwoCoins() {
	int coin1 free, coin2 free;
	if (coin1 == 0)
		return false;
	else if (coin2 == 0)
		throw Muli.fail();
	else
		return true; }
\end{lstlisting}
  \end{minipage}
\end{figure}

Non-deterministic execution is useful for applications involving search, i.\,e., an application would usually cause the runtime environment to evaluate more than one branch.
To that end, the runtime environment systematically evaluates multiple alternative branches in sequence.
Non-deterministic branching dynamically creates an implicit search tree that represents the various execution paths that lead to alternative outcomes of a program.
The goal of the present work is to make this search tree explicit at runtime.
It encodes the various execution paths of a program, the choices encountered along every path, and every path's outcome (i.\,e., solution or failure).
As there can be paths of infinite length, our search tree representation is non-strict.
Our search tree then serves as a basis for structured traversal by arbitrary search algorithms, including iterative deepening depth-first search.
Furthermore, by making the search tree explicit, it is possible to inspect the search tree at any given point in time, e.\,g., after search or even at an intermediate stage.
This way, the search tree aids in effective debugging.

This paper provides the following contributions:
\vspace{-9pt}\begin{itemize}
	\item A general \textit{search tree structure} for constraint-logic object-oriented programming that encapsulates execution state
	(\Cref{sec:searchtrees}).
	\item \textit{Search algorithm implementations} that traverse the search tree structure for finding solutions to constraint-logic object-oriented programs (\Cref{sec:searchstrategies}).
	\item A discussion of the implications of our work for executing object-oriented (imperative) programs non-deterministically (\Cref{sec:discussion}).
\end{itemize}

First of all, \Cref{sec:muli} introduces concepts of constraint-logic object-oriented programming, followed by an outline of the Muli virtual machine in \Cref{sec:mlvm}.

\section{Constraint-logic Object-oriented Programming} \label{sec:muli}

Constraint-logic object-oriented programming combines the flexibility of imperative and object-oriented programming with features from constraint-logic programming, namely logic variables, constraints, and search.
Muli is a constraint-logic object-oriented programming language that is based on Java~\cite{Dageforde2019cola}.

In Muli, \textit{logic variables} are declared in a way that is similar to declaring regular variables. As indicated in \Cref{lst:simplest-coin}, \par
	\lstinline|int coin free;| \par\hspace*{-\parindent}%
declares a logic variable of a primitive (integer) type.
Instead of assigning a constant value, the \lstinline|free| keyword specifies that \lstinline|coin| is a logic variable.
A logic integer variable can be used interchangeably with other integer variables, i.\,e.,
they can become part of conditions or arithmetic expressions and can be passed to methods as parameters~\cite{Dageforde2018semantics}.
In contrast to regular variables, logic variables are used symbolically.
Recent work is looking into support for reference-type logic variables \cite{Dageforde2019}, but here
we focus on logic variables of primitive types.

\textit{Constraints} are defined as relational expressions, (typically) involving logic variables.
For simplicity, Muli does not provide a dedicated language feature for imposing constraints.
Instead, a constraint is imposed whenever the flow of execution branches, such as when a branching condition is evaluated. Therefore, constraints are derived from boolean expressions.
For instance, in \Cref{lst:simplest-coin} \par
	\lstinline|if (coin == 0) {| $s_1$ \lstinline|} else {| $s_2$ \lstinline|}|\par\hspace*{-\parindent}%
\lstinline|coin| occurs in the condition and is not sufficiently constrained, so that the condition can be evaluated to either true or false.
As a result, the evaluation of the condition creates a \textit{choice}, from which alternatives are evaluated non-de\-ter\-min\-istic\-ally.
The runtime environment selects an alternative by imposing the corresponding constraint.
In our example, by imposing $coin \neq 0$ the runtime environment can proceed with the evaluation of $s_2$.
The runtime environment is supported by a constraint solver that is used for solving as well as for 
cutting execution branches early if their constraint system is inconsistent.

\textit{Search} transparently performs non-deterministic evaluation in combination with backtracking until a solution is found.
Implicitly, following a sequence of choices (and taking decisions at each choice) produces a (conceptual) search tree that represents the order of execution.
In such a search tree, inner nodes are choices and leaves represent alternative ends of execution paths.
In Muli, an execution path ends with a \textit{solution} (specified by either \lstinline|return| or \lstinline|throw|) or with a \textit{failure}, e.\,g., if a path's constraint system is inconsistent.
The full listing of our example in \Cref{lst:simplest-coin} demonstrates how solutions are returned.
After search completes, solutions of the example are \lstinline|false| and \lstinline|true| (in any given order).

Moreover, applications sometimes require an \textit{explicit failure} denoting the end of an execution path without a solution.
In Muli, an explicit failure is expressed by \lstinline|throw Muli.fail()|. Nevertheless, executing that statement will not return an exception. Instead, the statement is specifically interpreted by the runtime environment, resulting in backtracking.
\Cref{lst:two-coins} provides a slightly extended search region with three execution paths, one of which ends in a failure.

The main program is executed deterministically, whereas
all non-deterministic search is \textit{encapsulated}. This gives application developers control over search.
In addition to 
 coarse-grained control (i.\,e., requesting either a single solution or an array comprising all solutions),
 Muli offers
 fine-grained control by returning a Java stream that evaluates solutions non-strictly.
\lstinline|Muli.muli()| accepts a \lstinline|Supplier| and returns a stream of \lstinline|Solution| objects.
In Java (and, therefore, in Muli), a \lstinline|Supplier| denotes either a lambda expression or a method reference (both without arguments).
We refer to the method that is passed as an argument as a \textit{search region}, as it will be executed non-deterministically and therefore describes the constraint-logic object-oriented problem.
 Following the principles of the Java Stream API, solutions can be retrieved from the stream individually on demand~\cite{Dagefoerde2019encapstraversal}.
For instance, considering \Cref{lst:simplest-coin}, search is initiated by \par
\lstinline|Stream<Solution<Boolean>> stream = Muli.muli(self::flipCoin)|.

\section{Muli Logic Virtual Machine} \label{sec:mlvm}

The Muli Logic Virtual Machine (MLVM) is a runtime environment for Muli.
The MLVM is a custom Java Virtual Machine (JVM) that complies with the JVM Specification (see \cite{jvms8}) for deterministic execution
and adds modifications that support Muli-specific extensions,
particularly symbolic execution and non-deterministic execution  \cite{Dageforde2019cola}.
As in a regular JVM, execution state is represented in the MLVM by a combination of \textit{program counter (PC)}, a \textit{heap}, a stack of executed method frames (\textit{frame stack}), and an \textit{operand stack} per frame.
Additional state serves the purpose of supporting non-deterministic execution and constraints.
In particular, this includes the constraint stack and the trail.

The \textit{constraint stack} maintains the active constraint system,
i.\,e., the conjunction of all constraints on the stack \cite{Dageforde2019cola}.
Representing the constraint system in a stack structure is beneficial as constraints are added dynamically during execution.
Consequently, on backtracking, only the most recently added constraints need to be removed from the stack.
Moreover, the \textit{trail} records changes that are made to the virtual machine (VM) state during execution.
On backtracking, the information on the trail can be used to revert to a previous execution state.
More precisely, using the trail, backtracking achieves the specific state of the choice at which the next decision can be made.
In fact, the trail is therefore split up into incremental trails, one per choice, each describing how to backtrack towards the next choice.
In addition, in order to be able to not only backtrack to a choice (upwards along a search tree) but to achieve an arbitrary previous state (including downward navigation), the MLVM
maintains two trails per choice, one being the inverse of the other~\cite{Dagefoerde2019encapstraversal}.
In the following, we call the trail for backtracking \textit{backward trail}, as opposed to the \textit{forward trail} that is used to navigate downwards.

Like a regular JVM, the MLVM reads applications from bytecode and executes bytecode instead of the original source.
Muli's bytecode format is compatible with that described in \cite{jvms8}, merely adding custom attributes in order to represent logic variables \cite{Dageforde2019cola}.
For instance, the example application from \Cref{lst:two-coins} compiles to the bytecode instructions in \Cref{lst:two-coins-bytecode}.
Some bytecode instructions exhibit non-deterministic behaviour.
For instance, \lstinline|if_icmpne| in \Cref{lst:two-coins-bytecode} jumps to the specified instruction if the two integer operands on the operand stack are not equal. Otherwise, execution continues linearly with the following instruction.
If one or both operands are logic variables, both \textit{jumping} and \textit{not jumping} are feasible alternatives.
As logic variables are used in the current example,
the execution of \lstinline|if_icmpne| instructions creates choice points that offer two decision alternatives.
While \lstinline|if| instructions always provide two alternatives (i.\,e., jumping to the \lstinline|else| branch or not),
\lstinline|switch| instructions result in alternatives according to the number of \lstinline|case|s plus one for the \lstinline|default| case, each jumping to instructions accordingly.
\Cref{tab:nd-instructions} provides a reference of instructions that may exhibit non-deterministic behaviour and counts the decision alternatives from which the MLVM chooses.

\begin{lstlisting}[caption={Bytecode generated by the Muli compiler for the program in \Cref{lst:two-coins}.}, label={lst:two-coins-bytecode},float=t!,escapechar=@,xleftmargin=1.5ex]
0: iload_1             // coin1
1: iconst_0
2: if_icmpne @
\tikz[remember picture] \coordinate (a);
            @7@
\tikz[remember picture] \coordinate (b);
             @         // coin1 != 0
5: iconst_0
6: ireturn             // return false
@
\tikz[remember picture] \coordinate (abtarget);
@7: iload_2             // coin2
8: iconst_0
9: if_icmpne @
\tikz[remember picture] \coordinate (c);
            @16@
\tikz[remember picture] \coordinate (d);
              @        // coin2 != 0
12: invokestatic  #91  // fail()
15: athrow
@
\tikz[remember picture] \coordinate (cdtarget);
@16: iconst_1
17: ireturn            // return true@
\begin{tikzpicture}[remember picture, overlay, every node/.style={inner sep=0.5pt}]
\coordinate(aplus) at ($(b)+(0,1.5ex)$);
\coordinate(cplus) at ($(c)+(0,1.5ex)$);
\node[draw,circle, fit=(a)(b)(aplus)] (ab) {};
\node[draw,circle, fit=(c)(d)(cplus)] (cd) {};
\draw [-latex'] (ab.south west) -- ($(ab.south west -| abtarget) - (1.5ex,0)$) -- ($(abtarget) - (1.5ex,-0.7ex)$) -- ($(abtarget) - (0,-0.7ex)$);
\draw [-latex'] (cd.south west) -- ($(cd.south west -| cdtarget) - (1.5ex,0)$) -- ($(cdtarget) - (1.5ex,-0.7ex)$) -- ($(cdtarget) - (0,-0.7ex)$);
\end{tikzpicture}
@
\end{lstlisting}

\begin{table}[bt]
	\caption{Bytecode instructions that may cause non-deterministic branching upon execution. \lstinline|<cond>| is a placeholder for specific comparisons, e.\,g., \lstinline|eq| for equality.}
	\label{tab:nd-instructions}
	\centering
	\scalebox{0.9}{
\begin{tabular}{l@{~~}l@{~~}l}
	\toprule
	Triggering bytecode instruction                                            & Type of choice                  & No. of decisions \\
	\midrule
	\lstinline|If<cond>|, \lstinline|If_icmp<cond>|                            & \lstinline|if| instruction, integer comp. & 2                   \\
	\lstinline|FCmpg|, \lstinline|FCmpl|, \lstinline|DCmpg|, \lstinline|DCmpl| & floating point comparison                 & 2                   \\
	\lstinline|LCmp|                                                           & long comparison                           & 3                   \\
	\lstinline|Lookupswitch|, \lstinline|Tableswitch|                          & \lstinline|switch| instruction            & 1 per case + 1      \\ \bottomrule
\end{tabular}
}
\end{table}

Executing a bytecode instruction with non-deterministic branching creates a choice point in the MLVM~\cite{Dageforde2019cola}.
Prior to this work, the implementation of the choice point itself was responsible for managing the execution of its branches.
More specifically, executing a bytecode instruction created a choice point representation in the MLVM.
Consequently, the created choice point contained information about applicable branches, but also implemented the behaviour of search.
That is, upon creation, the choice point representation immediately selected the first decision alternative and applied it, thus committing to a specific branch.
The created choice point representations are stored in a stack of choice points.
The MLVM referred to the choice point stack during backtracking.
Starting from the top, it popped choice points until  reaching one with an alternative that had not been evaluated yet.
It then immediately committed to this alternative by adding its constraint and following its path.\footnote{Provided that the constraint system was still consistent.
Otherwise, backtracking occurred until the next choice point that offered an unevaluated, feasible alternative.}
As a consequence, the runtime environment never actually stored an explicit representation of the search tree.
Instead, the choice point stack merely maintained a single path through the (implicit) search tree.
Therefore, diverting from the currently executed path was not possible, effectively restricting the search capabilities of the MLVM to depth-first search.
All things considered, the previous MLVM used a complex, tangled mixture of responsibilities in which  bytecode-instruction implementations, choice point implementations, and the VM realise non-deterministic search in combination.

In a cleaner architecture,
\vspace{-9pt}\begin{itemize}
\item  executing a bytecode instruction declaratively creates choice objects and just returns them to the MLVM (instead of performing a decision right away) and
\item choice objects only hold information about available decision alternatives (but no implementation for taking decisions).
\end{itemize}
As a consequence, the MLVM is the only element that is allowed to change execution state by committing to decisions, instead of sharing this permission with choice objects or instruction implementations.
The search tree structure that we discuss subsequently facilitates an explicit representation that holds a declarative representation of choices and of the alternatives that each choice provides.
This serves as a clean basis for following arbitrary execution paths through the tree.

\section{Search Trees} \label{sec:searchtrees}

A declarative, explicit search tree representation lays the groundwork for following arbitrary execution paths instead of limiting execution to depth-first search only.
We first explain the conceptual representation, outlining the intuition of the elements that constitute the search tree.
Afterwards, we describe how a search tree is constructed dynamically during the execution of a Muli application.
Last, we abstractly describe navigation through the search tree as the basis for search.

\subsection{Representation} \label{subsec:searchtreerepresentation}

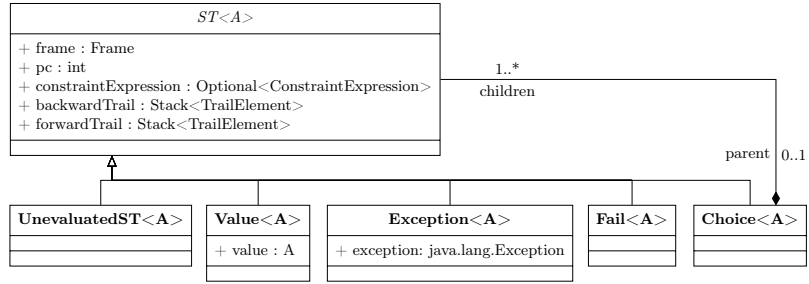
\begin{figure}[t]
  \centering
  \resizebox{.88\linewidth}{!}{\begin{tikzpicture}
  \tikzumlset{fill class=white}
  \umlclass[type=abstract,x=2.53,y=1,alias=ST,anchor=south]{ST<A>}{%
    + frame : Frame\\%
    + pc : int\\%
    + constraintExpression : Optional<ConstraintExpression>\\%
    + backwardTrail : Stack<TrailElement>\\%
    + forwardTrail : Stack<TrailElement>\\%
  }{}
  \umlclass[x=0,anchor=north,alias=UnevaluatedST]{UnevaluatedST<A>}{}{}
  \umlclass[x=3.2,anchor=north,alias=Value]{Value<A>}{+ value : A}{}
  \umlclass[x=7.1,anchor=north,alias=Exception]{Exception<A>}{+ exception: java.lang.Exception}{}
  \umlclass[x=10.8,anchor=north,alias=Fail]{Fail<A>}{}{}
  \umlclass[x=13.2,anchor=north,alias=Choice]{Choice<A>}{%
    }{}
  
  \coordinate(st) at (ST.-146);
  \coordinate(mid) at ($(st -| Fail.north)!.5!(Fail.north)$);
  
  \draw[tikzuml inherit style] (Fail.north) -- (Fail.north |- mid) -- (mid) -- (st |- mid) -- (st);
  \draw[tikzuml inherit style] (Value.north) -- (Value.north |- mid) -- (mid) -- (st |- mid) -- (st);
  \draw[tikzuml inherit style] (Exception.north) -- (Exception.north |- mid) -- (mid) -- (st |- mid) -- (st);
  \draw[tikzuml inherit style] (Choice.north) -- (Choice.north |- mid) -- (mid) -- (st |- mid) -- (st);
  \draw[tikzuml inherit style] (UnevaluatedST.north) -- (UnevaluatedST.north |- mid) -- (mid) -- (st |- mid) -- (st);
  
  \umlVHcompo[%
   attr2=parent|0..1, pos2=0.4, anchor1=50,
   attr1=children|1..*, pos1=1.8,
   ]{Choice}{ST}
\end{tikzpicture}}
  \vspace*{-1.5ex}
  \caption{Class diagram for the representation of search trees.}
  \label{fig:class-diagram-searchtree}
\end{figure}

Conceptually, our explicit search tree comprises five distinct node types.
There are node types for returned values, thrown exceptions, choices between non-deterministic branches, failed computations, and yet unevaluated search trees.
\Cref{fig:class-diagram-searchtree} shows a class diagram for our search tree representation.
Basically, this representation just corresponds to an algebraic data type and therefore does not implement any decision taking in contrast to the previously used choice points.

As solutions of a search region,
a \lstinline|Value| node holds the value returned by a computation
while an \lstinline|Exception| node does the same with an exception that has been thrown.
A \lstinline|Fail| node represents either an explicit failure or branches whose constraint system is inconsistent. As a consequence, it does not hold any values.
Furthermore, \lstinline|Choice| nodes store a list of subtrees which, in turn, reference their \lstinline|parent| choice.
Having an explicit reference to each node's parent allows for an easy and direct navigation through the search tree.
For the root node of a search tree, the \lstinline|parent| attribute is \lstinline|null|.
Finally, \lstinline|UnevaluatedST| serves as a proxy for subtrees that have not been evaluated yet,
facilitating non-strict usage.

Moreover, each node in the search tree stores fields that prepare for later execution.
The \lstinline|frame| and \lstinline|pc| fields represent a reference to the (mutable) stack frame and the value of the PC at which the node has been created.
Each node holds an optional constraint expression that has to be satisfied in order to reach this node, e.\,g., as a consequence of non-deterministic branching.
Additionally, the backward trail stores the changes to the VM state that were made in order to reach this node (thus preparing for backtracking),
whereas the forward trail stores changes that are needed in order to return to this node afterwards.
In combination, these fields
are used to properly manipulate the state of the MLVM during the traversal of the search tree, which is discussed in detail in \Cref{subsec:searchtreetraversal}.

\subsection{Construction} \label{subsec:searchtreeconstruction}

The actual search tree is constructed during search.
A search strategy is responsible for determining the order in which the search tree is traversed.
Regardless of the order, a search strategy evaluates \lstinline|UnevaluatedST| nodes as long as there are such nodes left and more solutions are demanded by the encapsulating program.
In general, the MLVM evaluates an \lstinline|UnevaluatedST| node by imposing the node's constraint and executing the bytecode of the search region starting from the PC, which the node points to, until either of the following situations occurs.
\vspace{-5pt}\begin{itemize}
\item The computation in the search region returns with a value,
\item an uncaught exception occurs during execution,
\item the method \lstinline|Muli.fail()| signals a failed computation, or
\item one of the instructions in \Cref{tab:nd-instructions} is executed, which results in the creation of a \lstinline|Choice| object.
\end{itemize}
\vspace{-5pt}
In any case, the \lstinline|UnevaluatedST| node in the search tree is replaced by its evaluated counterpart, i.\,e., by a \lstinline|Value|, \lstinline|Exception|, \lstinline|Fail|, or \lstinline|Choice| node.
Note that all children of a newly created \lstinline|Choice| node are unevaluated search trees initially.
Furthermore, state changes that were made during this evaluation are received from the MLVM and stored within the new node as its backward trail.

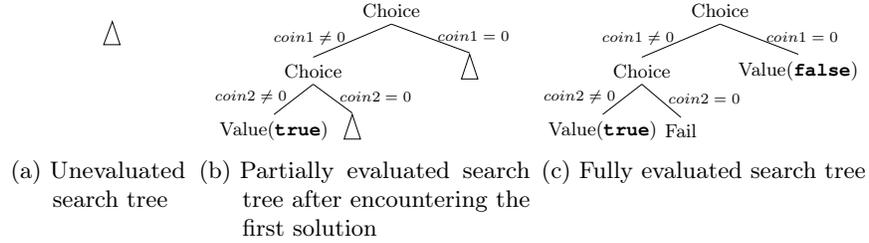
\begin{figure}[t]
  \centering
  \newcommand{\threesearchstagesscalefactor}{0.8}

\subfloat[Unevaluated search tree]{%
\centering
\hspace{1cm}
\scalebox{\threesearchstagesscalefactor}{
\begin{tikzpicture}[unevaluated/.style={every node,inner sep=0pt,yshift=0.7cm, xshift=-0.5cm},%
level/.style={level distance = 1cm}]
\node (st) [unevaluated] {}%
child { node{} edge from parent [opacity=0] child { node{} edge from parent [opacity=0] }}
;
\draw (st) -- ($(st) + (1ex, -3ex)$) -- ($(st) + (-1ex, -3ex)$) -- cycle;
\end{tikzpicture}
\hspace{1cm}
}
}
\subfloat[Partially evaluated search tree after encountering the first solution]{%
\centering	
\scalebox{\threesearchstagesscalefactor}{
\begin{tikzpicture}[unevaluated/.style={every node,inner sep=0pt,
yshift=2ex,},%
constraint/.style={yshift=1pt},%
level/.style={sibling distance=8em/#1, level distance = 1cm},%
unevaluatedtrail/.style={draw=ercisgrey,dashed}]
\node (st) {Choice}
child { node {Choice} 
	child { node {Value(\lstinline|true|)} edge from parent node[left,constraint] {\scriptsize $coin2 \neq 0$} }
	child { node [unevaluated]{} edge from parent node[right,constraint] {\scriptsize $coin2 = 0$} } 
	edge from parent node[left,constraint] {\scriptsize $coin1 \neq 0$}
}
child { node [unevaluated]{} edge from parent node[right,constraint] {\scriptsize $coin1 = 0$} };
\draw (st-2.center) -- ($(st-2.center) + (1ex, -3ex)$) -- ($(st-2.center) + (-1ex, -3ex)$) -- cycle;
\draw (st-1-2.center) -- ($(st-1-2.center) + (1ex, -3ex)$) -- ($(st-1-2.center) + (-1ex, -3ex)$) -- cycle;
\end{tikzpicture}
}
}
\subfloat[Fully evaluated search tree]{%
\centering	
\scalebox{\threesearchstagesscalefactor}{
\begin{tikzpicture}[unevaluated/.style={every node, draw},%
constraint/.style={yshift=1pt},%
level/.style={sibling distance=8em/#1, level distance = 1cm}]
\node {Choice}
child { node {Choice} 
	child { node {Value(\lstinline|true|)} edge from parent node[left,constraint] {\scriptsize $coin2 \neq 0$} }
	child { node {Fail} edge from parent node[right,constraint] {\scriptsize $coin2 = 0$} } 
	edge from parent node[left,constraint] {\scriptsize $coin1 \neq 0$}
}
child { node {Value(\lstinline|false|)} edge from parent node[right,constraint] {\scriptsize $coin1 = 0$} };
\end{tikzpicture}
}
}
  \caption{Different evaluation stages of the search tree corresponding to the search region in \Cref{lst:two-coins}. The constraint of each subtree is noted at the respective edge.}
  \label{fig:three-search-tree-stages}
\end{figure}

At the beginning of search, the search tree is unknown and therefore initially represented by a single \lstinline|UnevaluatedST| node.
The PC of that node points to the start of the search region and the optional constraint expression is left empty, since no constraints apply to the start of a search region.
Similarly, the trails are empty as this node has not yet been evaluated.
\Cref{fig:three-search-tree-stages} exemplarily shows three search trees for the program from \Cref{lst:two-coins} that all are evaluated to a different degree, and thus illustrate various intermediate evaluation stages that can occur during a search.
The illustration assumes a depth-first search strategy; therefore, other search strategies will result in different intermediate stages.

\subsection{Traversal} \label{subsec:searchtreetraversal}

The implementation of any search algorithm requires to be able to navigate through the search tree in any direction, i.\,e., upwards and downwards.
For example, if a branch of a search tree has been fully evaluated, search continues elsewhere.
While navigating through the search tree, it is important to ensure that the MLVM remains in a consistent state, which is what a node's forward and backward trail together with its frame and PC are used for.
In general, navigation takes place from an already evaluated node to another evaluated node, since only evaluated nodes have a trail (see \Cref{subsec:searchtreeconstruction}).
More specifically, a \lstinline|Choice| node is always the target node or source node when navigating upwards or downwards.

\begin{lstlisting}[caption={Methods for navigating upwards and downwards in a search tree.}, label={lst:searchtree-navigation},float=t]
void navigateUpwards(ST from, Choice to) {
  while (from != to) {
    if (from.constraintExpression.isPresent())
      constraintStack.pop();
    vm.processTrail(from.backwardTrail, from.forwardTrail);
    vm.setFrame(from.frame); vm.setPc(from.pc);
    from = from.parent; } }
void navigateDownwards(Choice from, ST to) {
  Stack<ST> nodes = new Stack<>();
  while (to != from) {
    nodes.put(to); to = to.parent; }
  while (!nodes.empty()) { to = nodes.pop();
    vm.setFrame(to.frame); vm.setPc(to.pc);
    vm.processTrail(to.forwardTrail, to.backwardTrail);
    if (to.constraintExpression.isPresent())
      constraintStack.push(to.constraintExpression.get()); } }
\end{lstlisting}

We navigate upwards in a search tree by following references to the parents until we reach the target node (e.\,g., the root), backtracking the VM state in the process.
In doing so, we remove previously imposed constraints from the constraint stack and undo the changes to VM state by processing the backward trails of nodes along the path.
At the same time, the backward trails are converted into forward trails so that a node from which we navigate away can be reached again later when navigating downwards, e.\,g., for the evaluation of another subtree of that node.
Last but not least, the frame and PC of the VM are set accordingly, using the information that was recorded at each node when it was created.

Navigating downwards is slightly more complicated as we first need to determine how to reach a target node from the current (source) node.
However, we always have a reference to the target.
Therefore, we can use the target's parents in order to find the path to the source.
Afterwards, we process the path in reverse order, thus getting from the source node to the target node.
We basically do the opposite of what is done in upwards navigation: For each node, we set the frame and PC to what is recorded in the node, apply the forward trail to reapply changes to the execution state, and impose a node's constraint if present.
Simultaneously to processing the forward trail, we convert it again into a backward trail to be later able to navigate upwards.
For clarity, \Cref{lst:searchtree-navigation} shows simplified implementations for navigating upwards and downwards, respectively.
Subsequently, these general navigation methods serve as primitives for traversal.

\section{Search Strategies} \label{sec:searchstrategies}

For the purpose of demonstrating how the explicit search tree representation can be employed for the implementation of search strategies, we outline the implementations of three concrete ones.

\subsubsection{Depth-first Search}

The implementation of depth-first search maintains a stack of unevaluated subtrees from the search tree.
At the beginning of the search, the initial node (see \Cref{subsec:searchtreeconstruction}) is pushed to the stack.
Then, depth-first search repeatedly pops an unevaluated search tree node from the stack and tries to evaluate it.
If its evaluation results in a \lstinline|Choice| node, its children are pushed to the stack and search continues by popping the next node from the stack (i.\,e., a local subtree).
Otherwise, if a \lstinline|Value| or \lstinline|Exception| node is encountered, the search strategy must be able to return the result to the encapsulating program. To that end, it reverts execution state to the state from the beginning of search using \lstinline|navigateUpwards|.
When search is picked up again, the search strategy uses \lstinline|navigateDownwards| in order to evaluate the next node from the stack.
Finally, if the node at hand is evaluated to a \lstinline|Fail| node, local backtracking is performed, i.\,e., we navigate upwards to the nearest parent that has at least one unevaluated subtree.

\subsubsection{Breadth-first Search}

Instead of a stack, a FIFO queue keeps track of unevaluated subtrees.
Beginning or resuming search dequeues nodes from the head of the queue.
In contrast, when a \lstinline|Choice| node is encountered, its children are enqueued at the end.
Another difference is the fact that breadth-first search requires navigating between arbitrary nodes within the search tree.
While it is of course possible to go over the root node, it is more efficient to navigate along a path going over the first common ancestor of the two involved nodes.
\Cref{lst:searchstrategy-findcommonancestor} shows a simple algorithm that determines the first common ancestor of two nodes in the search tree.
Once the first common ancestor is found, search combines \lstinline|navigateUpwards| (to the found ancestor) and \lstinline|navigateDownwards| in order to efficiently navigate between two arbitrary nodes.

\begin{lstlisting}[caption={Algorithm for finding the first common ancestor of two nodes.}, label={lst:searchstrategy-findcommonancestor},float=t]
Choice findCommonAncestor(ST a, ST b) {
  while (b != null) {
    add b to a set, b = b.parent; }
  while (!set.contains(a)) { a = a.parent; }
  return a; }
\end{lstlisting}

\subsubsection{Iterative Deepening Depth-first Search}

Our search tree can also be used to implement an interesting variant of iterative deepening search.
Iterative deepening provides the strength of depth-first search, while ensuring that solutions can be found even if there are execution paths of infinite length.
In iterative deepening, search is bounded by a constant maximum depth. Search proceeds in a depth-first manner until nodes are reached that are at the maximum depth.
In that case, search first evaluates other nodes up to that depth, thus assuming breadth-first search behaviour.
Only if additional solutions are required, search increases the bound, again by a constant, and so on.
In Muli, aided by the inverse trails, 
when the bound is increased, the runtime environment does not need to restart computation at the root which usually leads to reevaluation of known execution paths (and solutions).
Instead, it leverages the (partial) search tree and the recorded inverse trails in order to restart computation from known states that provide further alternatives.

\section{Discussion} \label{sec:discussion}

The implementation of our search tree structure in the MLVM facilitates the non-deterministic execution of imperative (object-oriented) programs in novel ways, using search strategies that could not be implemented without an explicit structure.
The existing depth-first search strategy has been reimplemented and is now based on the explicit search tree structure as well.
In order to ensure that the required changes do not adversely affect performance of depth-first search,
we first compare the runtime behaviour before discussing novel aspects of search.
Note that we measure only performance, not memory consumption. Obviously, maintaining the search tree requires more memory than merely storing the current execution path.
However, a possible memory optimisation would be to discard search tree nodes that belong to exhaustively evaluated subtrees --- especially in depth-first search strategies.

\begin{figure}[t]
	\vspace*{-3.5ex}
	\centering
	\includegraphics[width=0.35\linewidth]{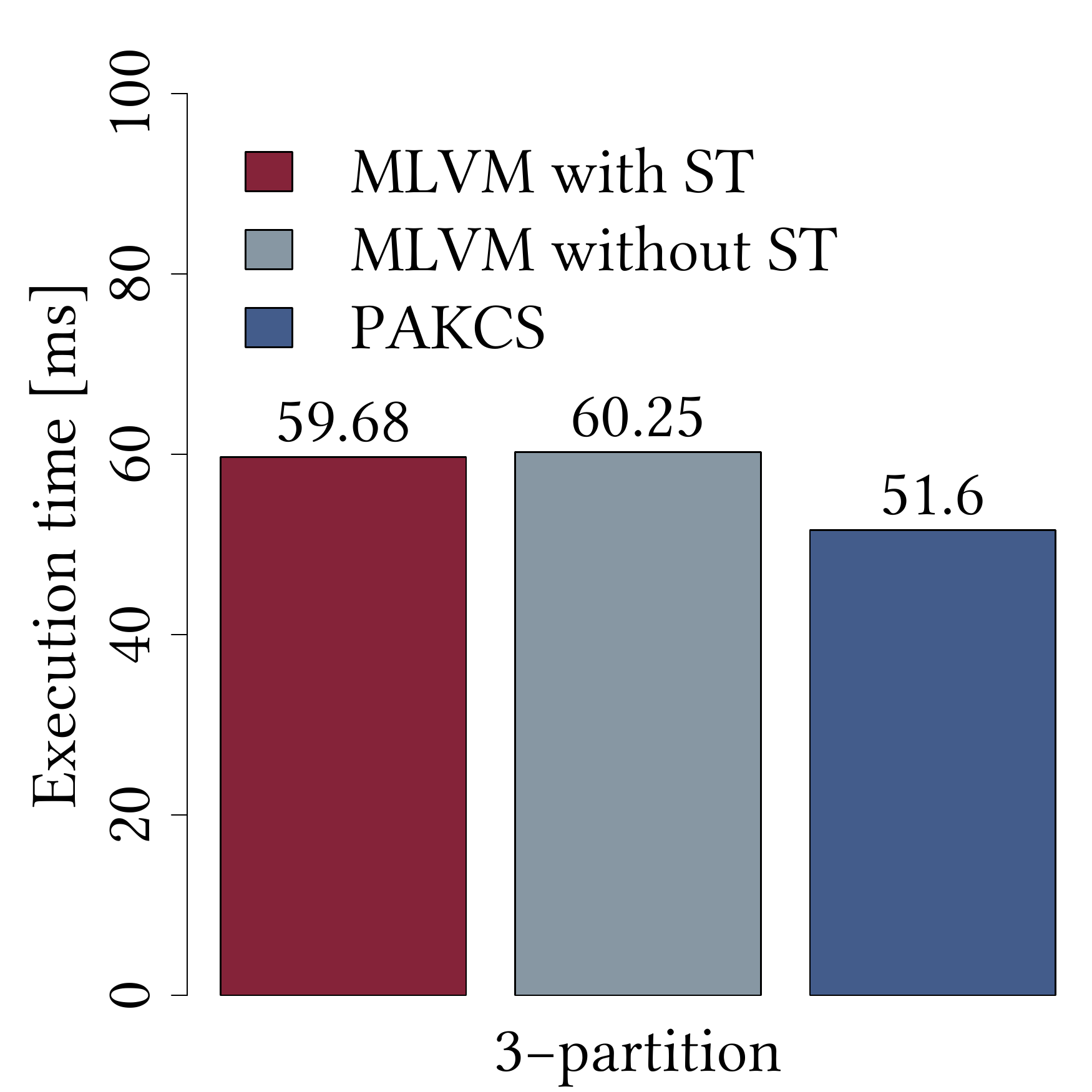}
	\hspace*{3em}
	\includegraphics[width=0.35\linewidth]{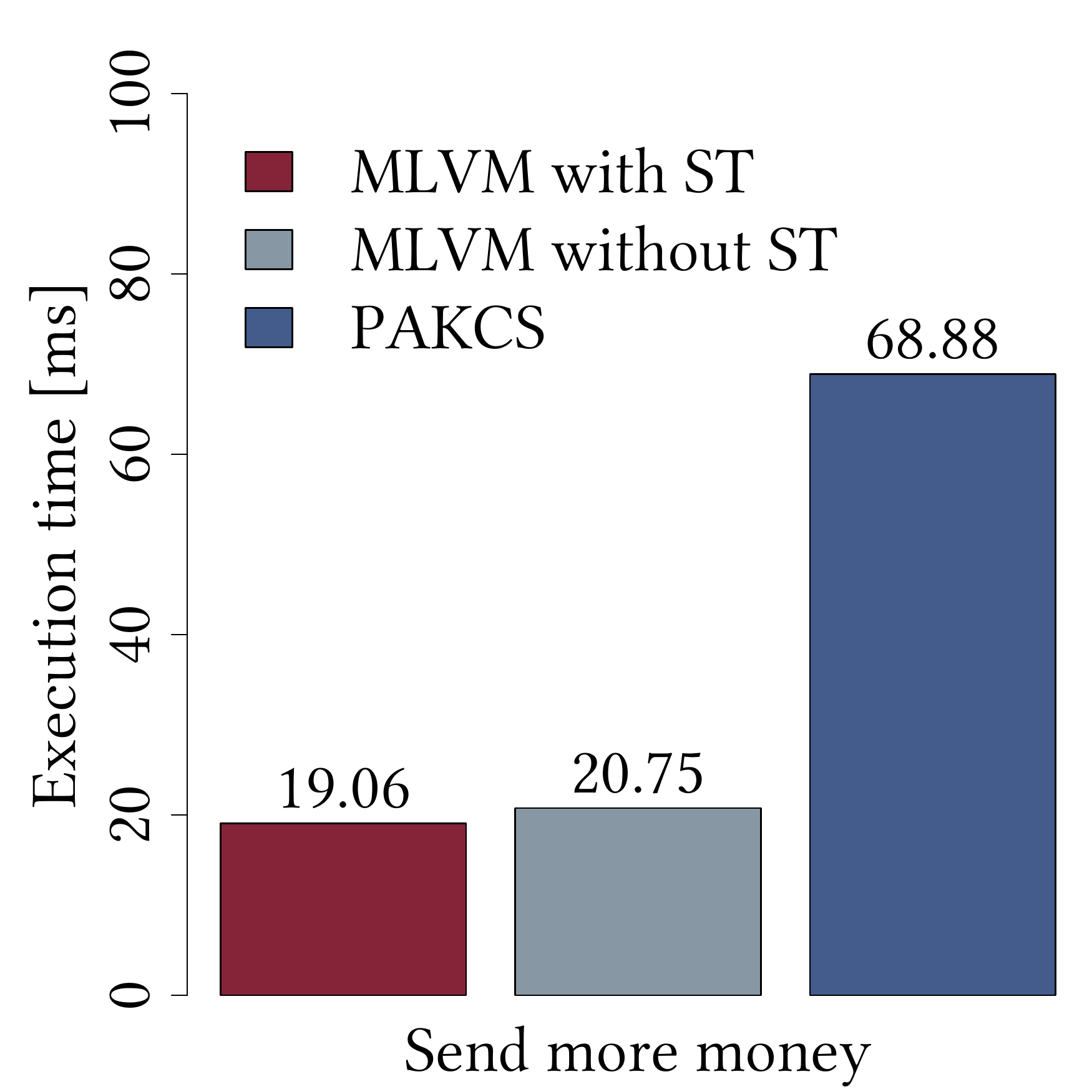}
	\vspace*{-2.5ex}
	\caption{Comparison of execution times in MLVM with or without explicit search trees, both using depth-first search. Execution times in PAKCS for reference.}
	\label{fig:dfscomparisons}
\end{figure}

We are interested in comparing the performance of depth-first search in the new search-tree-based and old choice-point-stack-based implementations.
To that end, a set of experiments is conducted in a modified MLVM that contains our search-tree structure as well as in an MLVM without modifications, each executed by OpenJDK 1.8.0\_212.\footnote{Ubuntu 18.04.2 with 4.15.0 x86\_64 GNU/Linux kernel; Intel Core i5-5200U CPU.}
Since the MLVM is executed by a JVM, we drop the first 15 executions in order to account for effects caused by just-in-time compilation and take the performance values of subsequent executions.
In total, we aggregate performance values of 500 executions per experiment, tackling classic search problems. The first experiment calculates a solution to the 3-partition problem for a fixed set of integer values using a depth-first search strategy. Until the first solution is found, search passes 374 choices.
The second finds a solution to the Send More Money puzzle. 
For reference, we also execute corresponding Curry implementations on PAKCS 2.1.1 using depth-first search.
\Cref{fig:dfscomparisons} features the average execution times.
Our experiment indicates that the implementation and use of an explicit search tree does not negatively affect depth-first search performance.
Moreover, the comparison to PAKCS is encouraging, seeing that Muli search regions offer competitive performance while providing support for using side-effects during non-deterministic execution.

Since the use of explicit search trees does not add visible overhead to execution times, we can focus on the benefits of using a search tree representation at runtime.
The MLVM now features additional search algorithms beyond depth-first search that all leverage the search tree structure.
In particular, using breadth-first search is novel to the non-deterministic execution of imperative programs that have side-effects.

Consider the search region from \Cref{lst:non-terminating-coin}.
For lack of a termination condition, there is one execution path that is infinite. Therefore, it is impossible to evaluate the search tree (or the application) strictly.
In our depth-first search implementation, the infinite execution path is the leftmost one.
As a result of this structure, depth-first search is unable to compute a single solution. 
In contrast, several solutions can be returned using a breadth-first or iterative deepening strategy, even though the tree can never be evaluated in full.
As a more sophisticated example, we have implemented a search region that finds solutions to the Water jugs problem.
Here the MLVM is unable to evaluate a full search tree as there are cyclic execution paths that result in valid solutions or failures.
We have executed these programs using the available strategies 500 times for up to ten seconds each and indicate the average number of solutions in~\Cref{tab:comparison-dfs-bfs}.

\begin{lstlisting}[caption={Muli search region featuring an infinite amount of execution paths.}, label={lst:non-terminating-coin},float=tb]
private static boolean nonTerminatingCoin() {
  int coin free;
  if (coin == 0)
    return true;
  else
    return nonTerminatingCoin(); }
\end{lstlisting}

\begin{table}[t]
	\caption{Comparison of search strategies w.\,r.\,t. the number of solutions that are returned within ten seconds.}
	\label{tab:comparison-dfs-bfs}
	\centering
	\scalebox{0.9}{
		\begin{tabular}{r@{~~~~~}r@{~~~}r@{~~~}r}
			\toprule
			& DFS & BFS & ID-DFS \\
			\midrule
			Simple infinite recursion & 0 & 1469.7 & 1555.2 \\
			Water jug problem & 0 & 29.5  & 34.4 \\
			\bottomrule
		\end{tabular}
	}
\end{table}

Note that the results do not imply that depth-first search is generally a bad strategy.
On the contrary, the combination of increased memory requirements  and the time needed for changing VM state using the trail still speaks against using breadth-first search by default.
Iterative deepening shares this disadvantage in case that additional levels of the search tree need to be evaluated (but is as efficient as depth-first search if the initial depth is sufficient).
Consequently, the results indicate that iterative deepening depth-first search is a good trade-off, if not a better strategy.
Further evidence is needed to conclusively argue that iterative deepening is a superior strategy in general.
In any case, both are useful strategies in certain situations in which depth-first search falls short.

The search tree structure that is presented in this paper is conceptually similar to the \lstinline|ST| structure known from the KiCS2 compiler for Curry~\cite{Hanus2012searchstrategies}.
However, Curry search trees only encode evaluation alternatives of an expression.
In contrast,
search trees for constraint-logic object-oriented programming need to encode the execution behaviour, i.\,e. VM state changes, that results from different alternatives.
Consequently, the state changes are recorded on the corresponding paths that lead to solutions, so that the VM can change state depending on the alternative that is being evaluated.
In our current work, we do this by maintaining the forward and backward trails on edges of the search tree.

Prior to our work, execution state of constraint-logic object-oriented programming in Muli was represented by the PC, frame stack, operand stacks, constraint stack, trail, and choice point stack.
Our work results in a slightly altered definition of execution state.
What previously was a choice point stack is now replaced by the search tree and a pointer to the current search tree node that is under evaluation.
In addition, a search algorithm is responsible for maintaining a suitable data structure that keeps track of the progress of traversing the search tree, e.\,g., a stack of not-yet-evaluated choices in depth-first search algorithms.

Moreover, the explicit search tree structure is useful for the development of constraint-logic object-oriented programs, as it can be helpful to visualise the structure of search.
Specifically, we can visualise at which points different kinds of choices are introduced and which solutions are encountered by the runtime environment.
During the development of the MLVM the search tree structure is useful for ensuring that non-deterministic branching and search algorithms are implemented correctly.
In contrast, the structure of the previous approach impeded the diagnosis of problems with non-deterministic execution, as only the current execution path was represented. Consequently, relevant information about previously encountered choices and solutions was lost, whereas this information is adequately represented in the explicit search tree.
All in all, the discussed benefits of an explicit search tree structure outweigh the increased memory requirements.

\section{Related Work} \label{sec:relatedwork}

For software testing, symbolic execution trees describe possible execution paths of an imperative program under test \cite{King1976,Majchrzak2009}.
Similar to our search tree, a symbolic execution tree represents choice points where execution branches and collects path constraints.
However, a symbolic execution tree usually describes the entire execution of an application.
In contrast, our search tree
for constraint-logic object-oriented programming 
describes the execution of specific application parts, namely the non-deterministic execution of a search region.
Its leaf node types are tailored to describing the result (i.\,e., solutions or failures) of execution paths.
Moreover, a symbolic execution tree is the result of performing depth-first search, whereas the dual trails of our search tree specifically supports arbitrary traversal.

The idea of using an explicit data structure for non-deterministic computations in order to facilitate different search strategies is extensively used in functional logic programming~\cite{Brassel2004encapsulating,Hanus2012searchstrategies}.
In functional logic programming, search trees cover non-determinism of expressions, i.\,e., they encode alternatives for the values that a pure expression can evaluate to.
In contrast to that, constraint-logic object-oriented programming is non-deterministic in its execution behaviour, which includes side-effects incurring during execution.
Therefore, the present search tree structure has to encode alternative behaviour, including side-effects, in addition to final results.
In addition to the representation usually used in functional logic programming, our representation includes node types for exceptions (as a different kind of solution) and unevaluated search trees.
The latter are a prerequisite for the on-demand construction of the search tree during search, which is innately given with the non-strict evaluation in functional logic programming.

An explicit data structure for representing a search tree structure has also been used in a monadic definition of constraint programming \cite{Schrijvers2009}.
In contrast to our work, it abstracts from side effects and asserts an ordering of subtrees.
Another explicit search tree is used for implementing a domain-specific language (DSL) for probabilistic programming in OCaml \cite{Kiselyov2009embedded}.
As OCaml is strict the on-demand characteristic of the search tree is modelled explicitly using lambda functions.
Although OCaml is not purely functional, the authors disregard backtracking w.\,r.\,t.\ behaviour, modelling only non-deterministic results of pure expressions.

As an alternative to using an explicit search tree, the interface of the probabilistic DSL in OCaml has also been implemented by using continuation passing style and by using delimited continuations, i.\,e., using \lstinline|shift| and \lstinline|reset| \cite{Danvy1990abstracting}.
Using continuations provides an implementation in direct style and removes the run-time overhead of the search tree data structure.
Therefore, implementing Muli by means of \lstinline|shift| and \lstinline|reset| is an interesting option for future work.
In this case, however, monadic reflection (i.\,e., inspecting the search tree) is expensive and its efficient implementation requires additional techniques \cite{Ploeg2015reflection}.

The concept of trails has initially been adapted from the trail described for the Warren Abstract Machine~(WAM)~\cite{Warren1983}
and has been extended towards dual trails for arbitrary execution state in~\cite{Dagefoerde2019encapstraversal}. Dual trails facilitate their use for backtracking upwards along a search tree  as well as for descending towards nodes that have been (partially) evaluated.
For their duality the two trails were originally termed trail and inverse trail.
Here we call them backward trail and forward trail, respectively, in order to improve clarity regarding the direction in which they are used.
Extending previous work, the present paper leverages dual trails for the implementations of search strategies other than depth-first search.

\section{Conclusion and Future Work} \label{sec:conclusion}

Our search tree structure represents the paths of non-deterministic execution of a search region.
A runtime environment of a constraint-logic object-oriented language can construct the search tree non-strictly while executing a search region,  thus
 encoding the solutions that are found as well as the execution behaviour of imperative code that leads to solutions or intermediate choices.
As a result, the explicit search tree representation can serve several purposes.
First, it provides a structure that arbitrary search strategies utilise for traversing the search tree.
Furthermore, we found it to make debugging of non-deterministic execution behaviour more effective by allowing developers who use a debugger to introspect intermediate state at breakpoints.
More opportunities for utilising the search tree in constraint-logic object-oriented programming will be part of future work.

We also extend Muli's runtime environment, the MLVM, to implement
 depth-first search, breadth-first search, and iterative deepening depth-first search.
Even though they are well-known as search algorithms for tree traversal, they are of special interest in the context of constraint-logic object-oriented programming where the search tree is not (fully) known before the program that it represents has been executed in its entirety.
The MLVM already supported depth-first search using the previous, unstructured approach, but our evaluation demonstrates that using a structured approach does not add any overhead.
On the contrary, the explicit representation provides opportunities for novel search algorithms that could not be used for executing constraint-logic object-oriented programs prior to our work.
The modifications have already been integrated into the open source MLVM and are available  at \url{https://github.com/wwu-pi/muli}.

The current work is the basis for future endeavours.
The search tree structure could be used for implementing an interactive search strategy in which a developer could manually decide how to explore the search space when a choice is encountered. This could be an additional aid for debugging.
Moreover, it is interesting to explore alternatives to explicit search trees, such as the use of delimited continuations for the implementation of non-deterministic execution.

\subsubsection*{Acknowledgements} 
The initial ideas that led to this work were conceived during the first author's visit to the University of Kiel.
The authors appreciate the valuable input of those that participated in the discussions; in particular, Sandra Dylus, Jan Christiansen, Jan Rasmus Tikovsky, and Michael Hanus.

%
%
%
\bibliographystyle{splncs04}
\bibliography{lit}

\begin{thebibliography}{10}
\providecommand{\url}[1]{\texttt{#1}}
\providecommand{\urlprefix}{URL }
\providecommand{\doi}[1]{https://doi.org/#1}

\bibitem{Brassel2004encapsulating}
Bra{\ss}el, B., Hanus, M., Huch, F.: Encapsulating {{Non}}-{{Determinism}} in
  {{Functional Logic Computations}}. Journal of Functional and Logic
  Programming p.~28 (2004)

\bibitem{Dageforde2019}
Dagef{\"{o}}rde, J.C.: {Reference Type Logic Variables in Constraint-Logic
  Object-Oriented Programming}. In: Silva, J. (ed.) WFLP 2018, pp. 131--144.
  Springer (2019). \doi{10.1007/978-3-030-16202-3\_8}

\bibitem{Dageforde2018semantics}
Dagef{\"{o}}rde, J.C., Kuchen, H.: {An Operational Semantics for
  Constraint-Logic Imperative Programming}. In: Seipel, D., Hanus, M., Abreu,
  S. (eds.) Declare 2017. pp. 64--80. Springer (2018).
  \doi{10.1007/978-3-030-00801-7\_5}

\bibitem{Dageforde2019cola}
Dagef{\"{o}}rde, J.C., Kuchen, H.: {A Compiler and Virtual Machine for
  Constraint-logic Object-oriented Programming with Muli}. Journal of Computer
  Languages  \textbf{53},  63--78 (2019). \doi{10.1016/j.cola.2019.05.001}

\bibitem{Dagefoerde2019encapstraversal}
Dagef{\"{o}}rde, J.C., Kuchen, H.: {Retrieval of Individual Solutions from
  Encapsulated Search with a Potentially Infinite Search Space}. In: Proc. 34th
  SAC. pp. 1552--1561. Limassol, Cyprus (2019). \doi{10.1145/3297280.3298912}

\bibitem{Danvy1990abstracting}
Danvy, O., Filinski, A.: Abstracting control. In: Proceedings of the 1990
  {{ACM}} Conf. on {{LISP}} and Functional Programming - {{LFP}} '90. pp.
  151--160. {ACM Press} (1990)

\bibitem{Hanus2012searchstrategies}
Hanus, M., Peem{\"o}ller, B., Reck, F.: Search strategies for functional logic
  programming. In: Proc. ATPS'12. pp. 61--74. GI LNI 199 (2012)

\bibitem{King1976}
King, J.C.: {Symbolic execution and program testing}. Communications of the ACM
   \textbf{19}(7),  385--394 (1976). \doi{10.1145/360248.360252}

\bibitem{Kiselyov2009embedded}
Kiselyov, O., Shan, C.c.: Embedded {{Probabilistic Programming}}. In: Taha,
  W.M. (ed.) Domain-{{Specific Languages}}, vol.~5658, pp. 360--384. {Springer}
  (2009)

\bibitem{jvms8}
Lindholm, T., Yellin, F., Bracha, G., Buckley, A.: {The Java{\textregistered}
  Virtual Machine Specification – Java SE 8 Edition} (2015),
  \url{https://docs.oracle.com/javase/specs/jvms/se8/jvms8.pdf}

\bibitem{Majchrzak2009}
Majchrzak, T.A., Kuchen, H.: {Automated Test Case Generation Based on Coverage
  Analysis}. In: TASE 2009. IEEE (2009). \doi{10.1109/TASE.2009.33}

\bibitem{Ploeg2015reflection}
van~der Ploeg, A., Kiselyov, O.: Reflection without remorse: Revealing a hidden
  sequence to speed up monadic reflection. ACM SIGPLAN Notices
  \textbf{49}(12),  133--144 (2015)

\bibitem{Schrijvers2009}
Schrijvers, T., Stuckey, P., Wadler, P.: {Monadic constraint programming}. JFP
  \textbf{19}(6),  663--697 (2009). \doi{10.1017/s0956796809990086}

\bibitem{Warren1983}
Warren, D.H.D.: {An Abstract Prolog Instruction Set}. Tech. rep., SRI
  International, Menlo Park (1983)

\end{thebibliography}

\end{document}